\documentstyle[pre,aps,floats,twocolumn,epsf]{revtex}

\begin{document}

\def\R{\mbox{$I\!\!R$}}

\def\A{\mbox{\bf A}}
\def\B{\mbox{\bf B}}
\def\D{\mbox{\bf D}}
\def\E{\mbox{\bf E}}
\def\F{\mbox{\bf F}}
\def\P{\mbox{\bf P}}
\def\U{\mbox{\bf U}}
\def\DF{\mbox{\bf DF}}
\def\DE{\mbox{\bf DE}}

\def\e{\mbox{$\hat{{\bf e}}$}}

\def\u{\mbox{\bf u}}
\def\w{\mbox{\bf w}}
\def\x{\mbox{\bf x}}
\def\y{\mbox{\bf y}}
\def\z{\mbox{\bf z}}

\def\zero{\mbox{\bf 0}}

\draft

\title{Designing coupling that guarantees synchronization
between identical chaotic systems}

\author{Reggie Brown}
\address{Physics Department and Department of Applied Science,
The College of William and Mary, Williamsburg, VA 23187-8795}

\author{Nikolai F. Rulkov}
\address{Institute for Nonlinear Science, University of
California, San Diego, La Jolla, CA 92093-0402}

\date{January 8, 1997}
\maketitle

\begin{abstract}
We examine synchronization between identical chaotic systems.
A rigorous criteria is presented which, if satisfied, guarantees 
that the coupling produces linearly stable synchronous motion.  The
criteria can also be used to design couplings that lead to stable
synchronous motion.  Analytical results from a dynamical system 
are presented.
\end{abstract}

\pacs{05.45.+b}

\narrowtext
Synchronization between chaotic systems has been the subject of
many theoretical papers over the last few years.  It has also
been experimentally observed in many 
systems~\cite{fy,pc,rnn,spie,hcp,uf,sushchik,sll,dg}.  Despite
these efforts many key issues remain open, and there are few
rigorous results that ensure the stability of synchronous
chaotic motion.  In most cases rigorous results are obtained 
using Lyapunov functions~\cite{hv,hmr,nr,wu}.  Unfortunately,
this method is not a regular approach and, in practice, it can
only be applied to particular examples.  The other rigorous 
approach is that of Ashwin~et.~al.~\cite{abs}.  To apply this
approach one must show that all normal Lyapunov exponents are 
negative for all measures of the dynamics.  For typical 
dynamical systems this leads to an intensive numerical analysis. 

There are a few special types of coupling between nonlinear
systems where rigirous analysis of the stability of synchronization
is straightforward.  One type is when the coupling transforms the
driven system into a stable linear system with time dependent 
driving.  A second is when the 
coupling is diagonal between all of the variables~\cite{fy}. 
In many practical cases these types of coupling can't be achieved. 
Thus, the state of the art does not give 
a practical answer to the following important question: Given an 
arbitrary dynamical system how can one {\em design} a physically 
available coupling scheme that is guaranteed to produce stable 
synchronous chaotic motion? 

This paper examines synchronization between identical systems 
with drive/response coupling.  The major result is a rigorous
criteria which, if satisfied, guarantees linearly stable 
synchronous motion.  More importantly, the criteria can be used
to {\em design} couplings that produce stable synchronized 
behavior.  The criteria only uses knowledge of the uncoupled
dynamics, and many of the important calculations can be performed
analytically.  Furthermore, the linearized stability equations
we examine arise in many other problems that have recently 
appeared in the literature.  A discussion of this last issue
is in our longer manuscript~\cite{br}.

Drive response synchronization between identical systems is 
modeled by
\begin{eqnarray}
\label{drive}
\frac{d \x}{dt} & = & \F(\x; t) \\
\label{response}
\frac{d \y}{dt} & = & \F(\y; t) + \E (\x - \y),
\end{eqnarray}
where \x\ is driving dynamics, \y\ is the response dynamics, 
and \E\ is a vector function representing the coupling.  For
these equations $\x$, $\y \in \R^d$ and $\E(\zero) = \zero$.
Synchronization occurs on an invariant manifold given by $\x 
=  \y$.  Obviously, if the coupling strength is below some
critical threshold then stable synchronous motion will not occur.
For some ($\F, \E$) pairs stable synchronous motion occurs only 
within a finite range of coupling strengths while for others
synchronization is never stable.

If one defines deviations from synchronization by $\w \equiv 
\y - \x$ then Eqs.~(\ref{drive}) and (\ref{response}) lead to
the following linearized equation for motion transverse to the 
synchronization manifold
\begin{equation}
\label{linear}
\frac{d \w}{dt} = \left[ \DF(\x) - \DE(\zero) \right] \w .
\end{equation}
In this equation $\DF(\x)$ is the Jacobian of \F\ evaluated
on the driving trajectory, \x, and $\DE(\zero)$ is the Jacobian
of \E\ evaluated at \zero.  The synchronization manifold is 
linearly stable if $\lim_{t \rightarrow \infty} \| \w(t) \| 
= 0$ for all possible driving trajectories $\x(t)$ associated 
with the chaotic attractor of the driving system.

To determine the behavior of $\w(t)$ in this limit divide
$\DF(\x) - \DE(\zero)$ into a time independent part, \A,
and an explicitly time dependent part, \B,
\begin{displaymath}
\DF(\x) - \DE(\zero) \equiv \A + \B(\x).
\end{displaymath}
(The nonuniqueness of this decomposition will be resolved
later.)  Assume \A\ can be diagonalized, transform to the
coordinate system defined by the eigenvectors of \A, and 
rewrite the linearized equations of motion as the following
integral equation~\cite{br}
\begin{eqnarray}
\z(t) & = & \U(t, t_0) \z(t_0) \nonumber \\
\label{int_eq}
 & + & \int_{t_0}^t \U(t,s) \left[ \P^{-1} \B[\x(s)] \P \right] 
\z(s) \, ds.
\end{eqnarray}
In this equation $\z \equiv \P^{-1} \w$ where $\P \equiv 
\left[ \e_1 \, \e_2 \, \cdots \e_d \right]$ and $\e_1, \, 
\e_2, \, \cdots \e_d$ are the eigenvectors of \A.  The 
ordering of the eigenvectors is given by the corresponding
eigenvalues, $\Re[\Lambda_1] \geq \Re[\Lambda_2] \geq \cdots
\geq \Re[\Lambda_d]$, where $\Re[\Lambda]$ is the real part
of $\Lambda$.  Also, $\U(t, t_0) \equiv \exp[\D (t - t_0)]$ 
is a time evolution operator, where $\D \equiv \P^{-1} \A \P$ 
is diagonal by assumption.

Linear stability of the synchronization manifold is determined
by $\| \z(t) \|$ in the $t \rightarrow \infty$ limit.  If one 
uses norms to convert Eq.~(\ref{int_eq}) into an inequality, 
and applies Gronwall's theorem, then one can define the following
decomposition~\cite{br}
\begin{eqnarray}
\label{def_A}
\A & = & \left\langle \DF \right\rangle - \DE  \\
\label{def_B}
\B & = & \DF - \left\langle \DF \right\rangle ,
\end{eqnarray}
where $\left\langle \bullet \right\rangle$ denotes a time average
along the driving trajectory.  In terms of this decomposition
the criteria for linear stability of synchronous motion is~\cite{br}
\begin{equation}
\label{cond}
- \Re[\Lambda_1] > \left\langle \| \P^{-1} \left[ \DF(\x) - 
\left\langle \DF \right\rangle \right] \P \| \right\rangle .
\end{equation}

Equations~(\ref{def_A})--(\ref{cond}) are our major results.
They represent definitions and conditions that indicate when
synchronous motion along a particular driving trajectory is {\em
guaranteed} to be stable to small perturbations in directions
transverse to the synchronization manifold.   The criterion is
rigorous and sufficient.  However, because it is based on norms
it is not necessary.  Indeed, numerical experiments indicate that
it tends to overestimate the necessary coupling strengths~\cite{br}.
Also, since the integral in Eq.~(\ref{cond}) is positive 
semi-definite the inequality can't be satisfied unless $\Re[
\Lambda_1] < 0$.  This condition is reminiscent of the discussion
of conditional Lyapunov exponents found in previous references.

The decomposition in Eqs.~(\ref{def_A}) and (\ref{def_B}) is optimal
in the sense that it minimizes the
right hand side of Eq.~(\ref{cond}).  We speculate that minimizing
this integral gives one the chance at satisfying the inequality. 
Furthermore, by inserting Eq.~(\ref{def_B}) into a Volterra expansion 
of Eq.~(\ref{int_eq}) one can show that, to second order, the 
criteria for linear stability is $\Re[\Lambda_1] < 0$~\cite{br}.
For any other decomposition this approximate stability criteria
will be correct to only first order.  Finally, for this decomposition 
Eq.~(\ref{cond}) reduces to $\Re[\Lambda_1] < 0$ for fixed points, 
a result that will not hold for other decompositions.

Equations~(\ref{def_A})--(\ref{cond}) depend explicitly on the 
measure of the driving trajectory.  Gupte and Amaritkar examined
synchronization using unstable periodic orbits as driving 
trajectories~\cite{ga}.  This, and later papers, show that, for
fixed coupling strength, different driving trajectories have 
different stabilities~\cite{abs,br,os}.  Recently, Hunt and
Ott~\cite{ho} numerically examined time averages on different
measures of a chaotic dynamical system and found that they
tend to assume their largest values on unstable periodic orbits
with the shortest periods.  This behavior is also discussed in 
Ref.~\cite{abs}.

Given these observations we conjecture that, in many practical 
cases, the unstable fixed points of the driving system will be 
the first measures on the synchronization manifold to go 
linearly unstable as the coupling strength is changed.  Thus, 
these trajectories should be the first measure to check for 
linear instability.  This conjecture is examined in our longer 
paper and is found to be true for the examples studied~\cite{br}.

Equation~(\ref{cond}) has a geometrical interpretation which can
be used to {\em design} couplings that yield stable synchronous 
motion.  The elements of $\DE(\zero)$ define a parameter space and
each side of Eq.~(\ref{cond}) defines a function in this parameter
space.  Thus, $\Sigma_{\bf x}$ and $\Sigma_{\Lambda}$, respectively
defined by $\left\langle \| \P^{-1} \left[ \DF(\x) - \left\langle
\DF \right\rangle \right] \P \| \right\rangle = {\rm const.} \equiv
C_1$ and $- \Re[\Lambda_1] = {\rm const.} \equiv C_2$, are families
of surfaces in this parameter space.  The boundary of the portion 
of the parameter space that yields linearly stable synchronization
is the intersection of these families of surfaces.  By choosing the
elements of $\DE(\zero)$ on portions of $\Sigma_\Lambda$ that are 
``above'' $\Sigma_{\bf x}$ one insures that the poles of \A\ are
sufficiently far into the left half plane to insure stability.
Thus, designing a coupling is similar to pole placement in control
theory~\cite{sontag}.

As an example we present an analysis of the following dynamical
system studied by Ott and Sommerer~\cite{os}
\begin{eqnarray}
\frac{d x}{dt} & = & v_x \nonumber \\
\frac{d v_x}{dt} & = & - \nu v_x + 4 x \left( 1 - x^2 \right) +
y^2 + f_0 \sin (\omega t) \nonumber \\
\label{os_ODE}
\frac{d y}{dt} & = & 2 v_y \\
\frac{d v_y}{dt} & = & - \nu v_y - 2 y \left( x - p \right) - 4k
y^3 \nonumber
\end{eqnarray}
where $\nu=0.05$, $f_0 = 2.3$, $\omega = 3.5$, $k=0.0075$ and
$p=-1.5$.  Originally, Ott and Sommerer examined the stability of
the invariant manifold defined by $y = v_y = 0$.  Their results 
indicate that for these parameter values motion on this manifold
is chaotic, the manifold itself is unstable, and only one stable
attracting set exists in $\R^4$.

As before, \x\ denotes the driving system and \y\ denotes the 
response system.  In principle the driving trajectory, $\x=[x, v_x,
y, v_y] \in \R^4$.  However, for this example we consider a 
driving trajectory restricted to the manifold examined by Ott and
Sommerer.  For this type of driving $\DF(\x)$ assumes a block 
diagonal form.  If we use block diagonal coupling then 
Eq.~(\ref{linear}) decomposes into motion parallel to, and 
perpendicular to the manifold examined by Ott and Sommerer.

For perpendicular motion
\begin{equation}
\label{two_dim}
\frac{d \w^{(\perp)}}{dt} = \left[ \DF^{(\perp)}(\x) -
\DE^{(\perp)} (\zero) \right] \w^{(\perp)},
\end{equation}
where
\begin{eqnarray*}
\DF^{(\perp)}(\x) & = & \left[ 
\begin{array}{cc}
0  &  1 \\
g^{(\perp)}(\x)  & -\nu
\end{array} \right]  \\
\DE^{(\perp)}(\zero) & = & \left[
\begin{array}{cc}
\epsilon^{(\perp)}_1 & \epsilon^{(\perp)}_4 \\
\epsilon^{(\perp)}_3 & \epsilon^{(\perp)}_2
\end{array} \right] ,
\end{eqnarray*}
and $g^{(\perp)}(\x) \equiv -2 (x - p)$.  (Equation~(\ref{two_dim})
is the same linear stability equation studied by Ott and Sommerer,
Eqs.~(7) and (8) in Ref.~\cite{os}).  An equation similar to 
Eq.~(\ref{two_dim}) involving $\w^{(\parallel)}$, $\DF^{(\parallel)}$,
$\DE^{(\parallel)}$, and $g^{(\parallel)}(\x) \equiv 4(1- 3x^2)$ exists
for motion parallel to the manifold.  (For the remainder of this letter
we drop the $\perp$ and $\parallel$ superscripts, and trust the reader
to perform calculations in {\em both} the perpendicular and parallel
subspaces.) 

It is easy to show that the eigenvalues of \A\ are
\begin{eqnarray}
\Lambda_\pm & = & \frac{- (\nu + \epsilon_1 + \epsilon_2)}{2} 
\nonumber \\
\label{os_Lambda_pm}
 & \pm & \frac{1}{2} \left[ (\nu + \epsilon_2 - \epsilon_1)^2 
+ 4 (1 - \epsilon_4) (\left\langle g \right\rangle - \epsilon_3)
\right]^{1/2}.
\end{eqnarray}
If $\Lambda_\pm$ are complex then $-\Re[\Lambda_1]$ can be made
arbitrarily large by increasing $\epsilon_1$ and/or $\epsilon_2$.
The case for real $\Lambda_\pm$ is more complicated, however, 
numerical results indicate that $-\Re[\Lambda_1]$ is maximized when
$\Lambda_\pm$ are complex~\cite{br}.  Because, $\left\langle \| 
\P^{-1} \left[ \DF(\x) - \left\langle \DF \right\rangle \right] \P 
\| \right\rangle$ diverges as $\Lambda_\pm$ transitions from real 
to complex $\epsilon$'s associated with this transition should to 
be avoided. These observations suggest that in order to satisfy the
condition for linear stability of the synchronization manifold one
should choose $\epsilon$'s so that $\Lambda_\pm$ are complex with 
imaginary parts that are not near zero.
\begin{figure}
\begin{center}
\leavevmode
\hbox{%
\epsfxsize=2.0in
\epsffile{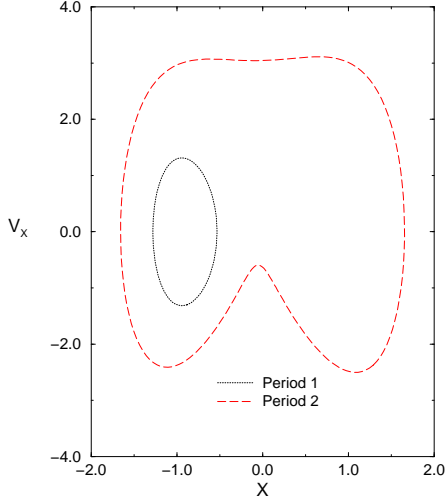}}
\end{center}
\caption{Period~1 and period~2 orbits of the Ott-Sommerer model.
\label{fig.orbits}}
\end{figure}

It is possible to show that if $\Lambda_\pm$ are complex then the
condition for linear stability of synchronous motion is
\begin{equation}
\label{cond4}
\nu + \epsilon_1 + \epsilon_2 > 4 C \left\langle | \Delta g | 
\right\rangle  ,
\end{equation}
where
\begin{equation}
\label{def_C}
C \equiv \left[ \frac{ - \left(1 - \epsilon_4 \right)^2}{(\nu +
\epsilon_2 -  \epsilon_1)^2 + 4 (1 - \epsilon_4) (\left\langle g
\right\rangle -  \epsilon_3)} \right]^{1/2},
\end{equation}
and $\Delta g(\x) \equiv g(\x) - \left\langle g \right\rangle$.
Equations~(\ref{os_Lambda_pm})--(\ref{def_C}), and the conjecture
that the $\epsilon$'s should be chosen so that $\Lambda_\pm$ are 
complex, is an {\em analytic solution} to the rigorous criteria
for synchronization.  Equating $C$ to a real positive
constant, in effect, selects surfaces from the families
$\Sigma_{\bf x}$ and $\Sigma_\Lambda$.  Each driving trajectory,
\x, corresponds to a different surface.

Since Eqs.~(\ref{os_ODE}) do not have fixed points we examined 
the SBR measure, and the measures associated with the periodic 
orbits shown in Fig.~\ref{fig.orbits} (the SBR measure is shown 
in Fig.~1 of Ref~\cite{os}).  Table~\ref{table.ave} shows 
numerically calculated values for $\left\langle g \right\rangle$
and $\left\langle | \Delta g | \right\rangle$.  
\begin{table}
\begin{tabular}{cccccc}
\multicolumn{2}{c}{Measure Type}  & $\left\langle g^{(\perp)}
\right\rangle$ & $\left\langle g^{(\parallel)} \right\rangle$ &
$\left\langle | \Delta g^{(\perp)} | \right\rangle$ &
$\left\langle | \Delta g^{(\parallel)} | \right\rangle$ \\
\tableline
\multicolumn{2}{c}{Period 1}  & -1.223 & -6.307 & 0.4769 &
5.142 \\
\multicolumn{2}{c}{Period 2}  & -3 & -7.767 & 1.678  & 10.30 \\
\multicolumn{2}{c}{SBR}  & -3 & -7.038 & 1.714  & 7.856 \\
\end{tabular}
\caption{\label{table.ave}}
\end{table}

We now explicitly examine several types of driving.  The first
is diagonal driving.  It is the first of the special cases where
rigirous results are straightforward~\cite{fy,hcp}.
Diagonal driving uses all components of \x\
and chooses $\epsilon_3 = \epsilon_4 = 0$, $\epsilon_1 = \epsilon_2
\equiv \epsilon$~\cite{fy,hcp}.  The parameter space is the
real line, \R.  For this type of driving $\Lambda_\pm$
are complex on each of our measures for all values of $\epsilon$.
Since $C$ is independent of $\epsilon$ its value is fixed, 
$\Sigma_{\bf x}$ is the entire parameter space, and $\Sigma_\Lambda$
is a family of points in \R.
Thus, for a particular driving trajectory the boundary for linear 
stability of the synchronization manifold (the intersection of 
$\Sigma_\Lambda$ with $\Sigma_{\x}$) is given by a point in \R.
The rigorous criteria for synchronization, Eq.~(\ref{cond4}), is
\begin{displaymath}
\epsilon > - \frac{\nu}{2} + 2 \left\langle | \Delta g |
\right\rangle \left[  \frac{-1}{\nu^2 + 4 \left\langle g
\right\rangle} \right]^{1/2}.
\end{displaymath}

Driving via postion uses only the position variables, $x$ and $y$.
The simplest example is $\epsilon_2 = \epsilon_3 = \epsilon_4 = 0$
and the parameter space is again \R.  It is useful to define new 
parameters $u \equiv \epsilon_1 + \nu$ and $w \equiv 1/C$.  In terms
of $u$ and $w$ Eqs.~(\ref{cond4}) and (\ref{def_C}) are
\begin{eqnarray*}
u w & > & 4 \left\langle | \Delta g | \right\rangle \\
- 4 \left\langle g \right\rangle  & = & (u - 2 \nu)^2 + w^2.
\end{eqnarray*}
If $\left\langle g \right\rangle < 0$ then these equations define a
hyperbola and a circle, respectively.  It is straightforward to show
that the circle does not intersect the hyperbola on the measures we
have examined.  Thus, for these driving trajectories the rigorous
condition for synchronization can not be satisfied.  (This does not
mean that stable synchronization {\em will not} result from this type 
of driving.  It only means that our analysis can not {\em guarantee}
that stable synchronization will result from this type of driving~\cite{br}.)

Another example of this type of driving uses the positions to drive both
the position {\em and} the velocity equations.  (This can sometimes
synchronize systems when simple driving via position does not produce
synchronization~\cite{od}.)  For this type of driving, $\epsilon_2 = 
\epsilon_4 = 0$ and the parameter space is $\R^2$.  Also, $\Lambda_\pm$
are complex for all $C > 0$, while Eqs.~(\ref{cond4}) and (\ref{def_C})
become
\begin{eqnarray*}
\epsilon_1 & > & - \nu + 4 C \left\langle | \Delta g | \right\rangle \\
\epsilon_3 & = & \frac{1}{4} \left( \nu - \epsilon_1 \right)^2 +
\left[ \left\langle g \right\rangle + \frac{1}{4 C^2} \right]. 
\end{eqnarray*}

These equations define a line and a parabola in $\R^2$.  For any
driving trajectory the line and the parabola are guaranteed to 
intersect.  Therefore, synchronization to {\em any} trajectory, \x,
is guaranteed to be linearly stable for coupling strengths on the 
parabola whose $\epsilon_1$ value is larger than the one 
associated with the intersection. 

Driving via velocity uses only the velocity variables, $v_x$ and
$v_y$.  As an example, let the velocities drive both the position
and the velocity equations.  Thus, $\epsilon_1 = \epsilon_3 = 0$, the
parameter space is $\R^2$, and $\Lambda_\pm$ are complex for $C>0$.
If we define new parameters $u \equiv \nu + \epsilon_2$ and $w 
\equiv (\epsilon_4 - 1)/C$ then Eqs.~(\ref{cond4}) and (\ref{def_C}) 
become 
\begin{eqnarray*}
u & > & 4 C \left\langle | \Delta g | \right\rangle \\
(2 C \left\langle g \right\rangle)^2 & = & u^2 + \left( 2 C
\left\langle g \right\rangle  - w \right)^2.
\end{eqnarray*}
These equations define a line and a circle, respectively.  It is
straightforward to show that the circle does not intersect the
line on the measures we have examined.  Therefore, the rigorous 
condition for synchronization can not be satisfied on these orbits.

In this paper we investigated the linear stability of the invariant 
manifold associated with synchronous behavior between coupled chaotic
systems.  Although we explicitly examined unidirectional coupling
our results are valid for bidirectional coupling, and for determining
the linear stability of invariant manifolds within a chaotic 
system~\cite{br}. 

Our major result is the rigorous criteria of 
Eqs.~(\ref{def_A})--(\ref{cond}).  When they are satisfied linear
stability of synchronous motion is guaranteed.  The criteria depends
on the measure of the driving dynamics and can yield different 
results for different driving trajectories.  The criteria can also
be used to design couplings that produce synchronization between 
coupled systems.

In closing this letter we discuss how noise and nonlinear effects
influence our results.  Assume the driving trajectory is a fixed
point, $\x_*$, and that Eq.~(\ref{cond}) is satisfied for $\epsilon > 
\epsilon_*$  (Equivalently, a period~1 orbit evaluated on a surface
of section.)  For this case the fixed point undergoes a co-dimension
one bifurcation (either pitchfork or transcritical) at $\epsilon = 
\epsilon_*$.  Linear stability analysis
does not take into account the unstable trajectories near $\x_*$ 
when $\epsilon \agt \epsilon_*$.  For arbitrarily small noise
amplitude there exists a range of $\epsilon$ values near $\epsilon_*$
where the noise will eventually push the response system beyond one
of the unstable orbits.  When this occurs the response system is 
forced to seek out an attracting state away from the synchronization
manifold.  Also, nonlinear effects could cause an unstable
orbit to approach $\x_*$ for some $\epsilon$ far from $\epsilon_*$.
If this occurs then small noise levels can also result in a loss of
synchronization.

R. Brown was supported by the Office of
Naval Research, grant No~N00014-95-1-0864 and the  AirForce
Office of Scientific Research, grant No.~F49620-95-1-0261. N.
Rulkov was supported the Department of Energy, grant 
No.~DE-FG03-95ER14516.


\begin{references}
\bibitem{fy} Fujisaka, H. and T. Yamada, {\em Prog. Theor. Phys.} 
{\bf 69}, 32 (1983).

\bibitem{pc} Pecora, L. M. and T. L. Carroll, {\em Phys. Rev.
Lett.}  {\bf 64}, 821 (1990); L. M. Pecora and T. L. Carroll,
{\em Phys. Rev.} {\bf 44A} , 2374 (1991).

\bibitem{rnn} Brown, R., N. F. Rulkov and N. B. Tufillaro, {\em Phys.
Rev.} {\bf 50E}, 4488 (1994).

\bibitem{spie} Pecora, L. M., {\em Chaos in Communications, SPIE 
Proceedings, San Diego, CA, 1993}, 2038, 2-25 (SPIE-The 
International Society for Optical Engineering,Bellingham, WA)

\bibitem{hcp} Heagy, J. F., T. L. Carroll and L. M. Pecora, {\em Phys.
Rev.} {\bf 50E}, 1874 (1994).

\bibitem{uf} Feldmann, U., {\em Synchronization of Chaotic Systems}
Ph.D. dissertation, Faculty of Electrical Engineering, Technical 
University of Dresden, July 1995.

\bibitem{sushchik} Sushchik, M. M., {\em Synchronized chaotic
oscillations}, Ph.D. dissertation, Physics Faculty, University
of California, San Diego, July 1996.

\bibitem{sll} de Sousa Vieira, M., A. L. Lichtenberg and M. A.
Lieberman, {\em Phys. Rev.} {\bf 46A}, R7359 (1992).

\bibitem{dg} Gauthier, D. J. and J. C. Bienfang, {\em Phys. Rev.
Lett.} {\bf 77}, 1751 (1996).

\bibitem{hv} He. R. and P. G. Vaidya, {\em Phys. Rev.} {\bf 46A},
7387 (1992).

\bibitem{hmr} Rodrigues, H. M., {\em Uniform ultimate boundedness
and synchronization} preprint CDSNS94-160 Georgia Institute of 
Technology.

\bibitem{nr} Rulkov, N. F.  et.al., {\em Int. J. Bif. Chaos} {\bf 2}, 
669 (1992).

\bibitem{wu} Wu, C. W and L. O. Chua, {\em Int. J. Bif. Chaos} 
{\bf 4}, 979 (1994).

\bibitem{abs} Ashwin, P., J. Buescu and I. Stewart, {\em Phys.
Letts.} {\bf 193A}, 126 (1994); P. Ashwin, J. Buescu and I. 
Stewart, {\em Nonlinearity} {\bf 9}, 703 (1996).

\bibitem{br} Brown, R. and N. F. Rulkov, ``Synchronization of chaotic
systems: Transverse stability of trajectories in invariant manifolds''
submitted for publication.

\bibitem{ga} Gupte, N. and R. E. Amritkar, {\em Phys. Rev.} {\bf
48E}, R1620 (1993).

\bibitem{os}  Ott, E. and J. C. Sommerer, {\em Phys. Lett.}
{\bf 188A}, 39 (1994).

\bibitem{ho} Hunt, B. R. and E. Ott, {\em Phys. Rev. Lett.} {\bf 76},
2254 (1996).

\bibitem{sontag} Sontag, E. D. {\em Mathematical Control Theory}
(Springer-Verlag, New York, 1990).

\bibitem{od} Ding, M.-Z. and E. Ott, {\em Phys. Rev.} {\bf 49E},
R945 (1994).
\end{references}
\end{document}